\RequirePackage{fix-cm}
\documentclass{article}
\usepackage[pdftex]{graphicx}
\usepackage{amsmath}
\usepackage{bm}
\usepackage{amssymb} 
\usepackage{mathtools} 
\usepackage{siunitx}
\usepackage{textgreek}
\usepackage{multirow}
\usepackage[normalem]{ulem}
\usepackage{longtable}
\usepackage{algorithmic}
\usepackage{algorithm2e}
\usepackage{natbib}
\usepackage{setspace}
\usepackage[nomarkers,figuresonly]{endfloat}
\usepackage{longtable}
\usepackage{arydshln}
\usepackage{pdflscape}
\usepackage{booktabs}
\usepackage{changepage}
\def\url#1{\expandafter\csname #1\endcsname}
\usepackage{enumerate}
\usepackage{url} 
\usepackage{setspace}
\usepackage{xspace}
\usepackage{multirow}
\usepackage[table,xcdraw]{xcolor}
\usepackage{adjustbox}
\usepackage{lineno,hyperref}
\modulolinenumbers[5]
\newcommand{\MATLAB}{\textsc{Matlab}\xspace}
\DeclareMathOperator*{\argmin}{argmin}

\newcommand{\eg}{e.g.\@ }
\newcommand{\ie}{i.e.\@ } 
\begin{document}

\title{Fast Computation of Highly $G$-Optimal Exact Designs via Particle Swarm Optimization 
}


\author{ Stephen J. Walsh and John J. Borkowski
}



\maketitle

\begin{abstract}
Computing proposed exact $G$-optimal designs for response surface models is a difficult computation that has received incremental improvements via algorithm development in the last two-decades. These optimal designs have not been considered widely in applications in part due to the difficulty and cost involved with computing them. Three primary algorithms for constructing exact $G$-optimal designs are presented in the literature: the coordinate exchange (CEXCH), a genetic algorithm (GA), and the relatively new $G$-optimal via $I_\lambda$-optimality algorithm ($G(I_\lambda)$-CEXCH) which was developed in part to address large computational cost. Particle swarm optimization (PSO) has achieved widespread use in many applications, but to date, its broad-scale success notwithstanding, has seen relatively few applications in optimal design problems. In this paper we develop an extension of PSO to adapt it to the optimal design problem. We then employ PSO to generate optimal designs for several scenarios covering $K = 1, 2, 3, 4, 5$ design factors, which are common experimental sizes in industrial experiments. We compare these results to all $G$-optimal designs published in last two decades of literature. Published $G$-optimal designs generated by GA for $K=1, 2, 3$ factors have stood unchallenged for 14 years. We demonstrate that PSO has found improved $G$-optimal designs for these scenarios, and it does this with comparable computational cost to the state-of-the-art algorithm $G(I_\lambda)$-CEXCH. Further, we show that PSO is able to produce equal or better $G$-optimal designs for $K= 4, 5$ factors than those currently known. These results suggest that PSO is superior to existing approaches for efficiently generating highly $G$-optimal designs.

\end{abstract}

\section{Introduction}
\label{sec:intro}
The groundbreaking publication which defined the optimal design concept predates R. A. Fisher's foundational 1935 work \emph{The Design of Experiments} \citep{atkRev, fisher:1935, firstGopt}. Kirstine Smith, a Danish female statistician working under Karl Pearson, published guidance in a 1918 Biometrika paper on how the residual error variance suggests designing an experiment used to fit a polynomial \citep{firstGopt}.  \citet{atkRev} note that this contribution is considered the seminal optimal design paper and was 30 years ahead of its time. This paper's contribution provided what we now refer to as  $G$-optimal designs for a $K=1$ factor experiment supporting a first up to sixth-degree polynomial. 

The exact $G$-optimal design generation problem is recognized as the notoriously difficult mini-max problem: we are seeking designs that minimize the maximum prediction variance over the study space. It wasn't until recent decades that modern computing resources were cheap enough to enable the application of well-suited meta-heuristic optimization approaches to this problem. \citet{jobo1} provided one of the earliest applications of Genetic Algorithms (GAs) for generating \emph{exact} optimal designs over the hypercube supporting a second-order response surface model in $K = 1, 2, 3$ design factors and for experiment sizes $N = 3, 4, 5, 6, 7, 8, 9$, $N = 6, 7, 8, 9, 10, 11, 12$, and $N = 10, 11, 12, 13, 14, 15, 16$ respectively. \citet{jobo1} adapted a GA to generate $G$-, $D$-, $I$-, and $A$-optimal designs and published a proposed catalog of exact optimal designs for each criterion. In contrast to continuous designs which can appeal to the \emph{General Equivalence Theorem} to verify the global optimality of a proposed continuous design, no such general theory exists for exact designs and so the community recognizes that proposed global exact optimal designs must be published and then challenged by future developments and algorithm applications. Therefore, the results provided in \cite{jobo1} have become a `ground truth' standard data set against which to compare results of newly developed algorithms to solve this problem. Until this point, the exact $G$-optimal designs of Borkowski (2003) remained the best-known designs for these scenarios and were subsequently reproduced by \cite{rodman} via an augmented application of the coordinate exchange and used as a benchmark by \cite{gIopt}. 

The next large investigation into generating and establishing the properties of exact $G$-optimal designs is provided by \cite{rodman}. In this paper, the authors adapted the coordinate exchange (CEXCH) algorithm of \cite{mncoord} in conjunction with Brent's minimization algorithm (used to score a candidate design on the $G$-scale) to generate exact $G$-optimal designs.  These authors covered all scenarios in Borkowski (2003) and were able to reproduce these designs with their new algorithm. A second contribution of this work was an extension of the algorithm to generate designs for a few $K = 4, 5$ factor cases---a design scenario not addressed to that point due to computation cost. The authors propose candidate exact $G$-optimal designs for $K = 4$, $N = 15, 20, 24$ and $K=5$, $N  = 21, 26, 30$ which support the second-order response surface model. A third contribution of this work was a detailed comparison of the resulting $G$-CEXCH generated exact $G$-optimal design's properties (via fraction of design space plots) with corresponding $D$- and $I$-optimal designs. The authors found that, for scenarios where post-hoc prediction was the primary objective, the $I$-optimal designs exhibited smaller prediction variance than the corresponding $G$-optimal designs  over large percentages of the design space (even though the $G$-optimal design has lower maximum prediction variance). This observation motivated the authors to recommend that the additional computing cost required for $G$-optimal designs may not be worth the effort and $I$-optimal designs might be preferred for many practical experimental scenarios.  Nonetheless, this catalog of proposed exact $G$-optimal designs for small response surface models remains a useful and referenced data set against which computational statisticians can compare the performance of the adaptation of new algorithms to this problem. 

The most recent publication on generating exact $G$-optimal designs is offered by \cite{gIopt}. The primary focus of this paper is the computational cost associated with generating $G$-optimal designs. These authors noticed a correspondence between continuous $I_\lambda$-optimal  designs and $G$-optimal designs. They propose a new algorithm that exploits the structural relationship between these two types of designs which first generates a continuous $I_\lambda$-optimal design as a starting point for the exact $G$-optimal design search, Then they apply the CEXCH algorithm to locally improve the candidate $G$-optimal design. We will refer to their algorithm as $G(I_\lambda)$-CEXCH. These authors are the first to completely run $G(I_\lambda)$-CEXCH and rerun the searches via GA by Borkowski (2003) in \MATLAB, and CEXCH by \cite{rodman} in JMP (they also reported the architecture of their PC) in order to compare computing efficiency of their proposed algorithm to the others. They implemented 200 runs of each algorithm for each design scenario discussed in \cite{jobo1} and also ran two new scenarios: $K = 4$, $N = 17$ and $K = 5$, $N = 23$ with only $G(I_\lambda)$-CEXCH (they indicated that $G$-CEXCH would have taken 25 and 166 days respectively to complete the task, and so this algorithm was not run on these scenarios). To briefly summarize their results, they found that, with respect to finding the highest $G$-optimal designs, the GA was superior (finding the best design each set of 200 algorithm runs) but that $G$-CEXCH and $G(I_\lambda)$-CEXCH produced designs with high relative efficiency vs. $GA$ generated designs with relative-efficiencies of 90\% or greater. However, the GA achieves this superiority at high computational cost exhibiting 2-orders of magnitude increase in the number of objective function evaluations as compared to the $G(I_\lambda)$-CEXCH. $G(I_\lambda)$-CEXCH was demonstrated to produce designs with high $G$-optimality as well as $G$-CEXCH but with computing times at a fraction of $G$-CEXCH for the higher dimensional searches. Thus they propose that $G(I_\lambda)$-CEXCH is a good choice for generating exact $G$-optimal designs as it is 1.) demonstrated to produce designs with high relative efficiency to GA generated designs and 2.) does this more efficiently than existing algorithms.

Particle swarm optimization (PSO) is a mature, simple, but highly effective meta-heuristic optimization algorithm (well suited for finding global optima on multimodal objectives) in the class of evolutionary algorithms (including GA) and has not been applied to this problem to date. Since its publishing by \cite{opso_1995} in the machine learning literature, this algorithm has attained widespread application and success in many scientific domains---to date Google tracks this paper with over 69,216 citations (queried January 17, 2022) which illustrates its widespread impact in algorithm development research and applications to engineering and applied science domains. This success notwithstanding, PSO has seen relatively few applications in optimal design, and for those publications that do address the topic, they tend to focus on using PSO to generate optimal \emph{continuous} designs---see, for example, \cite{chen0, wong, chen1, chen2, Chen3, chen4}. There have also been some applications of PSO, with apparent high success, outside of the response surface design framework---see, for example, \cite{lukemire} for an application of PSO to generating $D$-optimal designs for logistic-regression models and \cite{joseph} for an application of PSO to generating space filling designs. The contribution of our paper is: 1) a first application of PSO to the exact $G$-optimal design generation problem, 2) a review and summary of PSO generated results against those algorithms and results published over the last 20 years, and 3) we update the proposed catalog of exact $G$-optimal designs with new-found improved $G$-optimal designs for many of the studied scenarios. 

In the next section we present our notation and define the exact $G$-optimal design generation problem. In section 3 we develop a yet unpublished extension of PSO for generating exact optimal designs. In section 4 we compare $G$-PSO generated designs to those provided by \cite{jobo1}, \cite{rodman} and \cite{gIopt}. In section 5 we provide discussion, conclusions, and future research directions. Based on the new information reported in this paper, we conclude with a recommendation for which algorithm appears to be best suited for generating exact $G$-optimal designs.



\section{$G$-Optimal Designs}
\label{sec:gopt}
As in previous publications, we will be working with the second-order response surface model under standard assumptions. Our notation is as follows. Let $\mathcal{X}$ denote a design space. Without loss of generality with respect to generating optimal designs on irregular hyper-rectangular design regions, it is common, in a response surface setting, to assume that the design factors have been scaled to $[-1, 1]$. In which case the design space is the $K$-dimensional hypercube $\mathcal{X} = [-1, 1]^K$ for $K$ factors. Let $\mathbf{x}' \in \mathcal{X}$ denote a $1 \times K$ point in the design space. For $N$ specified design points, the set of distinct design points $\mathbf{x}'_i, \text{ } i = 1, 2, \hdots, N$ are organized as the rows of the $N \times K$ design matrix 
$$\mathbf{X} = \begin{pmatrix} \mathbf{x}'_1 \\ \mathbf{x}'_2 \\ \vdots \\ \mathbf{x}'_N \end{pmatrix}.$$
Next, we provide some notation that clarifies the difference between the design space and the model space---the latter of which is required to evaluate the properties and the quality of a candidate design $\mathbf{X}$.  Let $p$ represent the number of parameters of the model that one intends to fit to the data. We denote the $N \times p$ model matrix as $\mathbf{F}(\mathbf{X})$. Here we have emphasized that the model matrix is a function of the design matrix. For simplicity, however, from this point forward we will simply represent the model matrix by the shorthand $\mathbf{F}$. A similar device is required for a model vector, which is a function that maps a candidate design point to the model space. We adopt the notation $\mathbf{f}'(\mathbf{x}')$ which is a $1 \times p$ row-vector. 

We provide an example for the case of an $N$ point design with $K = 3$ design factors on the second order model to clarify the use of this notation. In this case there are 
$$p = {K + 2 \choose 2} = {5 \choose 2} = 10 $$
parameters in the second-order model \citep{rsm, jobo1}.  The design matrix for this model is
$$\mathbf{X} = \begin{pmatrix} \mathbf{x}'_1 \\ \mathbf{x}'_2 \\ \vdots \\ \mathbf{x}'_N \end{pmatrix} = \begin{pmatrix} x_{11} & x_{12} & x_{13} \\ x_{21} & x_{22} & x_{23}  \\ & \vdots & \\ x_{N1} & x_{N2} & x_{N3} \end{pmatrix}.$$ The second-order model, for a single design point $\mathbf{x}' = (x_1 \quad x_2 \quad x_3 )$ in scalar form is written
$$y = \beta_0 + \sum_{i = 1}^K\beta_ix_i + \sum_{i=1}^{K-1}\sum_{j = i + 1}^K\beta_{ij}x_i x_j + \sum_{i=1}^K \beta_{ii}x_i^2 + \epsilon, $$ 
and so the mapping of the design point $\mathbf{x}'$ into the model space is the $1 \times p$ vector
$$\mathbf{f}'(\mathbf{x}') = \begin{pmatrix} 1 & x_1 & x_2 & x_3 & x_1 x_2 & x_1 x_3 & x_2 x_3 & x_1^2 & x_2^2 & x_3^2 \end{pmatrix}.$$
The mapping of the design matrix into the model matrix is
$$ \mathbf{F}(\mathbf{X}) := \mathbf{F} = \begin{pmatrix} 1 & x_{11} & x_{12} & x_{13} & x_{11} x_{12} & x_{11} x_{13} & x_{12} x_{13}   & x_{11}^2 & x_{12}^2 & x_{13}^2 \\ 1 & x_{21} & x_{22} & x_{23} & x_{21} x_{22} & x_{21} x_{23} & x_{22} x_{23}   & x_{21}^2 & x_{22}^2 & x_{23}^2 \\ \vdots & & \vdots & & \vdots & & \vdots & &\vdots &  \\ 1 & x_{N1} & x_{N2} & x_{N3} & x_{N1} x_{N2} & x_{N1} x_{N3} & x_{N2} x_{N3}   & x_{N1}^2 & x_{N2}^2 & x_{N3}^2  \end{pmatrix}. $$ 
This enables writing the full second-order linear model in the $N$ design points in matrix form
$$\mathbf{y} = \mathbf{F}\bm{\beta} + \bm{\epsilon},$$
where $\bm{\beta}$ is a $p \times 1$ vector of parameters. Under the ordinary least squares (OLS) framework, it is assumed that $\bm{\epsilon} \sim \mathcal{N}_N(\mathbf{0}, \sigma^2 \mathbf{I}_N)$ where $\mathcal{N}_N$ denotes the $N$-dimensional multivariate normal distribution. The least-squares estimate of the $\bm{\beta}$'s (which in this framework are also ML estimates) is
$$\hat{\bm{\beta}} = (\mathbf{F}'\mathbf{F})^{-1}\mathbf{F}'\mathbf{y},$$
which have variance-covariance
$$\text{Var}(\hat{\bm{\beta}}) = \sigma^2 (\mathbf{F}'\mathbf{F})^{-1}.$$
A prediction of the response at new point $\mathbf{x}'$ is denoted as $\hat{y}(\mathbf{x}')$, and the variance of the mean predicted value is
$$\text{Var}(\hat{y}(\mathbf{x}')) = \sigma^2\mathbf{f}'(\mathbf{x}')(\mathbf{F}'\mathbf{F})^{-1}\mathbf{f}(\mathbf{x}').$$

The total information matrix for $\bm{\beta}$, specifically $\mathbf{M}(\mathbf{X}) = \mathbf{F}'\mathbf{F}$, plays an important role in optimal design of experiments---all optimal design objective functions are functions of this matrix.

While $\mathcal{X}$ denotes the space of candidate \emph{design points} $\mathbf{x}'$, a \emph{design matrix} $\mathbf{X}$ is a collection of $N$ such design points. Thus, the space of all candidate designs can be denoted
\begin{equation}
	\label{eq:exactModspace}
	\mathbf{X} \in \bigtimes_{j = 1}^N \mathcal{X} = \bigtimes_{j = 1}^N [-1, 1]^K = [-1, 1]^{NK} = \mathcal{X}^N,
\end{equation}
and the practitioner must choose a design from $\mathcal{X}^N$ to implement the experiment in practice. Note that the optimization problem has dimension $N*K$ (the dimension of the optimization search for the designs studied here is sometimes, erroneously, referred to as `small' and only viewed as a function of the number of design factors $K$). An optimality criterion is used to define which candidate designs $\mathbf{X} \in \mathcal{X}^N$ are `good' designs. An optimization algorithm (such as GA, $G$-CEXCH, or $G(I_\lambda)$-CEXCH) is used to search $\mathcal{X}^N$ to find the `best', or hopefully, globally optimal design. A $G$-optimal design is that design which minimizes the maximum scaled prediction variance over all points of prediction in $\mathcal{X}$. Recall that the variance of a mean prediction is
$$\text{Var}(\hat{y}(\mathbf{x}')) = \sigma^2\mathbf{f}'(\mathbf{x}')(\mathbf{F}'\mathbf{F})^{-1}\mathbf{f}(\mathbf{x}')$$ 
for any point of prediction $\mathbf{x}' \in \mathcal{X}$.  The scaled prediction variance (SPV) removes the scale parameter $\sigma^2$ and re-scales to $N$ by multiplying by the factor $N / \sigma^2$. Thus, SPV is defined as (for a candidate design $\mathbf{X}$)
\begin{align}
\label{eq:spvsc}
	\text{SPV}(\mathbf{x}') &:= \mathbf{f}'(\mathbf{x}') \mathbf{M}^{-1}(\mathbf{X}) \mathbf{f}(\mathbf{x}')\nonumber  \\
	& = N \mathbf{f}'(\mathbf{x}')(\mathbf{F}'\mathbf{F})^{-1}\mathbf{f}(\mathbf{x}').
\end{align}
Note that defining $SPV$ as in the second line of Eq. \ref{eq:spvsc} is necessary, as will be explained leading up to Eq. \ref{eq:gbound}.

The $G$-score of an arbitrary design $\mathbf{X}$ is defined as the maximum scaled prediction variance over all points of prediction $\mathbf{x}' \in \mathcal{X}$
\begin{align}
	\label{eq:gscore}
	G(\mathbf{X}) &:= \max_{\mathbf{x}'\in\mathcal{X}}  \text{SPV}(\mathbf{x}') \nonumber \\
	 &= \max_{\mathbf{x}'\in\mathcal{X}} N \mathbf{f}'(\mathbf{x}')(\mathbf{F}'\mathbf{F})^{-1}\mathbf{f}(\mathbf{x}').
\end{align}
Eq.\@ \ref{eq:gscore} indicates that for a fixed candidate design $\mathbf{X}$, scoring the candidate on the $G$ scale is itself an optimization problem. The $G$-optimal design $\mathbf{X}^*$ is that design which minimizes, over all designs $\mathbf{X} \in \mathcal{X}^N$, the maximum scaled prediction variance, namely
\begin{align}
	\label{eq:gdesign}
	\mathbf{X}^* & := \argmin_{\mathbf{X} \in \mathcal{X}^N} G(\mathbf{X}) \nonumber \\
	& = \argmin_{\mathbf{X} \in \mathcal{X}^N} \max_{\mathbf{x}'\in\mathcal{X}}  \text{SPV}(\mathbf{x}') \nonumber \\
	 &= \argmin_{\mathbf{X} \in \mathcal{X}^N} \max_{\mathbf{x}'\in\mathcal{X}} N \mathbf{f}'(\mathbf{x}')(\mathbf{F}'\mathbf{F})^{-1}\mathbf{f}(\mathbf{x}').
\end{align}
Eq.\@ \ref{eq:gdesign} shows that finding the $G$-optimal design is a minimax problem. This optimization has proved notoriously difficult to solve because neither of the optimizations required to compute $\mathbf{X}^*$ are convex. The scale of $G$ for an arbitrary design has known bounds.  The General Equivalence Theorem of \cite{kiefer} demonstrates that the lower bound on $G(\mathbf{X})$ is
\begin{equation} 
	\label{eq:gbound}
	G(\mathbf{X}) = \max_{\mathbf{x}'\in\mathcal{X}}  \text{SPV}(\mathbf{x}') \ge p.
\end{equation}
That is, the smallest that the maximum scaled prediction variance of candidate design $\mathbf{X}$ may be is $p$, the number of parameters. This apprently gives a way to verify if a proposed exact $G$-optimal design is globally optimal, however, not all design scenarios have globally $G$-optimal designs that will achieve this lower bound. Further, if design $\mathbf{X}^*$ is globally $G$-optimal and achieves the result in Eq. \ref{eq:gbound}, then for this design $\text{SPV}(\mathbf{x'}) = p$ at all diagonals of the hat matrix $\mathbf{F}(\mathbf{F}'\mathbf{F})^{-1}\mathbf{F}'$ and
$\text{SPV}(\mathbf{x'}) \le p$ at all other points of prediction $\mathbf{x}' \in \mathcal{X}$ \citep{rsm}. These are the only results from the theory of continuous optimal designs that also apply directly to exact optimal designs. The result stated in Eq. \ref{eq:gbound} is an important result---it means that the scale of $G$-optimality is not arbitrary (at least up to the number of parameters in the model $p$). It is customary to exploit this fact and score candidate designs on the $G$-efficiency scale,
\begin{equation}
	G_{\text{eff}}(\mathbf{X}) = 100 \times \frac{p}{G(\mathbf{X})},
\end{equation}
in order to gauge the quality of a candidate design $\mathbf{X}$ (larger $G_{\text{eff}}$ on this scale implies a better design). Last, the \emph{relative efficiency} of two candidate designs may be computed as
\begin{equation}
	\label{eq:grelleff}
	G_{\text{releff}}(\mathbf{X}_{1}, \mathbf{X}_{2}) = 100\frac{G_{\text{eff}}(\mathbf{X}_{1})}{G_{\text{eff}}(\mathbf{X}_{2})}
\end{equation}

The reason for constructing an algorithm to search $\mathcal{X}^N$ globally for the $G$-optimal design is that once one has a candidate design $\mathbf{X}$, one must gauge its quality via Eq. \ref{eq:gscore}. In the context of the larger search of $\mathcal{X}^N$ for the globally optimal design $\mathbf{X}$, each candidate $\mathbf{X}$ must be scored on the $G$-scale via Eq. \ref{eq:gscore}, and this is a separate optimization of a multimodal function. There are two published approaches in the literature suggesting ways to score a candidate matrix $\mathbf{X}$ via computing or approximating $G(\mathbf{X})$.

First, authors \cite{jobo1} and \cite{gIopt} exploit the symmetry of the SPV-surface for a $G$-optimal design and recommend using a  $5^K$ point grid over $\mathcal{X}$ with each factor containing $x_j \in \{-1, 0.5, 0, 0.5, 1\}$ for $j = 1, \hdots, K$  grid-points. The full grid defined as $\mathbf{G}_\mathcal{X} = \{-1, 0.5, 0, 0.5, 1\}^K$ \citep{gIopt}. Once one has a candidate design, the SPV is evaluated on $\mathbf{G}_\mathcal{X}$ and the maximum value is taken as an approximation of the $G$-score for the candidate design. While this approach computes an approximate $G$-score for a candidate design $\mathbf{X}$ both authors have noted that, due to the symmetry of the $G$-surface for $G$-optimal designs, the approximation is quite adequate and supports finding highly $G$-optimal designs.

A second approach is published by authors \cite{rodman} who apply a different approach to scoring a candidate design on the $G$-scale during the evolution of the $G$-CEXCH algorithm. In this algorithm once a candidate is available (e.g. after every coordinate exchange) they employ Brent's minimization algorithm to search for the maximum prediction variance for the candidate design over 

Regarding the choice of method for scoring a candidate design on the $G$-scale in this paper: we implemented the $5^K$ grid approach recommended by \cite{jobo1} and \cite{gIopt} for several reasons. First, it is easy to implement in the context of the larger $G$-optimal design search on $\mathcal{X}^N$. Second, in preliminary studies we also found these approximations to the $G$-score to be adequate and subject to small error. Third, in preliminary studies, we attempted a separate optimization approach similar to \cite{rodman} but found that if a candidate design was mis-scored the effect was disastrous on the evolution of the algorithm searching $\mathcal{X}^N$ for the optimal design.

In the next section we describe the PSO algorithm and provide an extension of PSO to generate $G$-optimal designs used to produce the results presented in this paper.

\section{Particle Swarm Optimization}
\label{sec:meth}
PSO has been demonstrated to perform very well for high-dimensional optimization of multimodal objective functions, see for example, \cite{opso_1995, spso1_2007, spso2_2011, hal_spso, opso2_1995, ken_1997, eberhart_1998, clerc2, engel, zombook, bratton}. PSOs strengths include: 1) few-to-no assumptions about the properties of the objective function $f$ to be optimized, 2) PSO is demonstrated to be strongly robust to entrapment in local optima, and thereby is a good `match' to the exact optimal design generation problem, 3) simplicity---the core function of the algorithm can be explained via two simple update equations, and 4) in contrast to other meta-heuristics where studying a range of tuning parameters can yield more efficient searches for specific problems, PSO only has three tuning parameters and these have been studied extensively theoretically and empirically with optimal values demonstrated for searches (such as ours) that fall on the typical Euclidean geometry, see \cite{clerc2, ebershi_3, clerc1, shi2} among others. Therefore, we employ the values of the PSO tuning parameters as suggested in the machine learning literature, and these values are guaranteed to 1) ensure a large degree of swarm exploration of the search space and 2) ensure that the particles in the swarm eventually settle on a consensus point for the optimal solution to the optimization problem \citep{clerc2, hal_spso}.

PSO is suited to perform the calculation
\begin{equation}
 \mathbf{x}^* = \argmin_{\mathbf{x} \in \mathcal{X}} f(\mathbf{x})
\end{equation}
for objective $f$ with $K\times 1$ input vector $\mathbf{x}$. The algorithm proceeds by first randomly drawing several candidate solution points (\ie the particles) $\mathbf{x}_i$ according to a multivariate uniform distribution in $\mathcal{X}$ for $i = 1, \hdots, S$ particles. Similarly each particle is assigned an initial velocity $\mathbf{v}_1$ which governs particle $i$'s step size and direction for its search of the optimizer of $f$. Then, for iteration $t$ of the PSO algorithm the particles step through the search space (searching for the optimum of $f$) according to velocity and position update equations
\begin{flalign}
  && \mathbf{v}_i(t+1) & = \omega \mathbf{v_i}(t) && \text{(inertia)}   \label{eq:inertia} \\ 
	&&  + & c_1 U_K(\mathbf{0}, \mathbf{1}) \odot\left(\mathbf{p}_{\text{best},i} - \mathbf{x}_i(t) \right) && \text{(cognitive)} \label{eq:cognitive}  \\ 
	&&  + & c_2U_K(\mathbf{0}, \mathbf{1}) \odot\left(\mathbf{g}_{\text{best}} - \mathbf{x}_i(t) \right) && \text{(social)} \label{eq:social} \\
	&& \mathbf{x}_i(t+1) & = \mathbf{x}_i(t) + \mathbf{v}_i(t+1) && \text{(position update)}\label{eq:posup}
\end{flalign}
where $\odot$ represents the Hadamard product (elementwise multiplication), $\mathbf{p}_{\text{best},i}$ is the `personal best' position in $\mathcal{X}$ regarding fitness on $f$ that particle $i$ has visited to time $t$ (\eg  this is the particle's `memory'), $\mathbf{g}_{\text{best}}$ represents the `group best' position in $\mathcal{X}$ regarding fitness on $f$ that particle $i$ found by the swarm to time $t$ (\eg  this the the swarm's `communication'), and $U_K(\mathbf{0}, \mathbf{1}) = \{u_j\}$ for $j = 1, \hdots, K$ where $u_j \overset{i.i.d.}{\sim} U(0,1)$ is a random scaling factor for each velocity update component. Finding optimal velocity weighting constants $\omega, c_1, $ and $c_2$ was the subject of intense research over approximately a decade. In this work we employ, as recommended by the extensive PSO literature for optimization problems of our nature, $ \omega = 0.72984, c_1 = c_2 = 2.05\omega = 1.496172 $. These values are guaranteed to balance particle exploration while ensuring that the velocities eventually scale down so that the swarm radius eventually decreases to 0, \ie the particles in the swarm come to a consensus regarding a solution to $\argmin_{\mathbf{x} \in \mathcal{X}} f(\mathbf{x})$ \citep{hal_spso, clerc1, clerc2}. Last, the personal best position of particle $i$, $\mathbf{p}_{\text{best},i}$, and the group's best position, $\mathbf{g}_{\text{best}}$, are updated at each iteration $t$ via the simple logic
\begin{align*}
		& \textbf{if} f(\mathbf{x}_i) < f(\mathbf{p}_{\text{best},i}) \\ 
		& \hspace{0.4cm}\mathbf{p}_{\text{best},i} \leftarrow \mathbf{x}_i \\
		& \hspace{0.4cm}\textbf{if} f(\mathbf{p}_{\text{best},i}) < f(\mathbf{g}_{\text{best}}) \\
	  & \hspace{0.8cm}\mathbf{g}_{\text{best}} \leftarrow \mathbf{p}_{\text{best},i}. \\
\end{align*}
The use of $\mathbf{g}_{\text{best}}$ in the description of PSO above represents the swarm \emph{communication topology}---in this case all members of the swarm communicate and share knowledge of the global value $\mathbf{g}_{\text{best}}$ and so this is referred to as the global communication topology \citep{hal_spso, engel}. The most effective communication topology was heavily researched during the development of PSO. Several other communication topologies were considered. A random local topology, where communication neighborhoods are randomly instantiated for particle $i$ with a small number, say 3, of expected connections (\ie particle $i$ can only share the information its gained about $f$ to the particles in its neighborhood) was found to be effective in improving the robustness of the swarm to entrapment in local optima for multimodal objectives and exact optimal design problems \citep{engel, walsh}. In this approach if the swarm does not find a better fitness on $f$ at time $t$ then the communication links are randomly re-instantiated at time $t+1$, thus creating a new local communication neighborhood for particle $i$ and, hence, new information sharing among the particles. In this scenario the term $\mathbf{l}_{\text{best},i}$, which represents the best position in $\mathcal{X}$ found \emph{in the communication neighborhood of particle $i$} to time $t$, replaces $\mathbf{g}_{\text{best}}$. Thus, the local communication topology encourages the swarm to explore $\mathcal{X}$ more thoroughly at the expense of more iterations. However, we view this as an acceptable tradeoff if we are seeking the global optima of a difficult-to-optimize objective function. 

\subsection{Extending PSO to the Optimal Design Problem}
PSO is formulated to optimize a function $f$ which takes a $K\times 1$ vector input $\mathbf{x}$. In contrast, optimal design requires optimizing an objective function which takes an $N \times K$ dimensional matrix input $\mathbf{X}$. We considered using the standard formulation of PSO by implementing $\mathbf{x}^* = \text{vec}(\mathbf{X})$ where vec is the vectorization operator (\ie it stacks the columns of a matrix) as this would give us access to off-the-shelf implementations of PSO. Due to programming and research considerations we implemented an alternate but equivalent approach by extending the PSO update equations as
\begin{flalign}
  && \mathbf{V}_i(t+1) & = \omega \mathbf{V_i}(t) && \text{(inertia)}   \label{eq:ch3matinertia} \\ 
	&&  + & c_1 \mathbf{U} \odot\left(\mathbf{P}_{\text{best},i} - \mathbf{X}_i(t) \right) && \text{(cognitive)} \label{eq:ch3matcognitive}  \\ 
	&&  + & c_2\mathbf{U} \odot\left(\mathbf{L}_{\text{best}_i} - \mathbf{X}_i(t) \right) && \text{(social)} \label{eq:ch3matsocial} \\
	&& \mathbf{X}_i(t+1) & =  \mathbf{X}_i(t) + \mathbf{V}_i(t+1) && \text{(position update)}\label{eq:ch3matposup}
\end{flalign}
where the $N\times K$ matrix $\mathbf{U} = \{u_{nk}\}$ with $u_{nk} \overset{i.i.d.}{\sim} U(0, 1)$ (\ie all elements are randomly drawn from a uniform distribution). It is evident that equations (\ref{eq:ch3matinertia}-\ref{eq:ch3matposup}) reduce to Eq.s (\ref{eq:inertia}-\ref{eq:posup})  when $N=1$ and the resulting vectors are transposed. Thus, in our formulation, a particle is now a candidate design matrix $\mathbf{X}_i$ and has a corresponding velocity matrix $\mathbf{V}_i$ which governs how particle matrix $i$ will step through the search space $\mathbf{X} \in \mathcal{X}^N$ in search of the exact globally optimal design. To fully implement PSO to search for optimal designs on the hypercube, there are several other technical problems to address (\eg how do we deal with design points that leave $\mathcal{X}$ during the search?). For these we have adapted standard PSO solutions as found in the literature \citep{hal_spso}. Our full PSO algorithm for generating optimal designs on the hypercube is provided in Algorithm \ref{alg:bpso}. The user can specify lower and upper bounds on each of the $K$ factors, and in line 1 we have set defaults as $\mathbf{l}_b = -\mathbf{1}_K$ and $\mathbf{u}_b = \mathbf{1}_K$ respectively without loss of general study because the initialization of PSO takes into account the size of the search space (a type of standardization), but these may be relaxed to irregular hyper-rectangular regions if required for alternate searches. Lines 4-11 show the initialization of the particle swarm (which is here a set of randomly drawn design matrices and initial velocities). In line 4 the $j$th row of initial design matrix is drawn corresponding to a uniform distribution with lower and upper bounds $\mathbf{l}_b$ and $\mathbf{u}_b$ \citep{hal_spso}. The $j$th row of the velocity matrix of particle $i$ is initialized similarly as shown in line 5. Line 6 shows that the stepsize of elements of the design matrices is controlled for each of the $k$th factors via parameter $\nu_k^{\text{max}}$, and \cite{hal_spso} recommends setting this parameter to half the length of the search space in one dimension though they do indicate that the use of this parameter is superfluous with the use of the values of $\omega, c_1, c_2$ as employed in this paper (because the initialization of PSO takes into account the size of the search space which is a type of standardization). Line 7 indicates the instantiation of $\mathbf{N}_i$ which represents a vector indicating which particles are in the communication neighborhood of particle $i$. Further, this vector is created by subroutine \texttt{genNeighbors} according to \cite{hal_spso} which sets neighborhood links for particle $i$ randomly such that the expected number of communication links is 3 but may range from 1 (\ie the particle only informs itself) to $S$ (\ie the particle can inform all other particles). With initial particle positions and velocities determined, line 10 initializes the solution, and line 11 shows initialization of the neighborhood best locations. Line 13 starts the swarm search which iterates until a stopping criteria is met. Lines 14-16 check the solution at time $t$ and if it is unchanged from time $t-1$ then the communication neighborhoods, $\mathbf{N}_i$, are randomly re-instantiated. Lines 18-29 describe velocity updating, position updating, and swarm knowledge updating as previously discussed in this section. Note the subroutine \texttt{confine} on line 22. It is possible that a particle moves outside the search space during the search and so we implement a `reflecting wall' particle confinement as in \cite{hal_spso}. Function \texttt{confine} checks every element of $\mathbf{X}$. If any element has moved out of the search space, that position is set to the closest point on the boundary (in its respective dimension) of the $\mathcal{X}$ and its velocity is halved and reversed in that direction (i.e. it `bounces off the wall'), see \cite{walsh} for technical details. Last, lines 31-24 show how the local neighborhood best locations are updated after the personal best's $\mathbf{P}_{\text{best},i}$ are updated. The algorithm progresses in this manner until a stopping criterion is met. 

\onehalfspace
\begin{algorithm}[htbp]
\caption{PSO for Generating Exact Optimal Designs on the Hypercube}
\label{alg:bpso}
  \begin{algorithmic}[1]
		\vspace{0.2cm}
    \STATE \textbf{Input:} Objective function $f$, number of particles $S$, search space bounds $\mathbf{l}_b = -\mathbf{1}_K$ and $\mathbf{u}_b= \mathbf{1}_K$
		\vspace{0.2cm}
		\STATE \hspace{0pt}{\textbf{//} Randomly draw $S$ candidate designs $\mathbf{X}_i$ and initialize }
		\STATE \hspace{0pt}{\textbf{for each }{$i = 1, \ldots, S$ \textbf{do}}} \label{initialize:loop:1}
		\STATE \hspace{8pt}{$\mathbf{x}'_{ij} \sim U_K(\mathbf{l}_b, \mathbf{u}_b) \text{ for } j = 1, \hdots, N$  giving candidate design $\mathbf{X}_i$ with rows $\mathbf{x}'_{ij}$ }
		\STATE \hspace{8pt}{$\mathbf{v}'_{ij} \sim U_K\left(\frac{\mathbf{l}_b - \mathbf{x}_{ij}}{2}, \frac{\mathbf{u}_b - \mathbf{x}_{ij}}{2} \right)\text{ for } j = 1, \hdots, N$  giving velocity matrix $\mathbf{V}_i$ with rows $\mathbf{v}'_{ij}$ }
		\STATE \hspace{8pt}{$\{v_{ijk}\leftarrow\min\{v_{ijk}, v_{k}^\text{max}\}\}$ for $j = 1, \hdots, N$ \textbf{//} limit stepsize}
		\STATE \hspace{8pt}{$\mathbf{N}_i \leftarrow \texttt{genNeighbors}(\mathbf{X}_i)$ \textbf{//} generate communication neighborhood for particle $i$}
		\STATE \hspace{8pt}{$\mathbf{P}_{\text{best},i} \leftarrow \mathbf{X}_i$ \textbf{//}set initial personal best position}
		\STATE \hspace{0pt}\textbf{endfor}
		\vspace{0.2cm}
		\STATE \hspace{0pt}{$\mathbf{G}_{\text{best}} \leftarrow \underset{\mathbf{X}_i \in \{\mathbf{X}_1, \mathbf{X}_2, \ldots, \mathbf{X}_S\}}{\mathrm{argmin}} f(\mathbf{X}_i)$ \textbf{//} current best design among swarm}
		\vspace{0.2cm}
	  \STATE \hspace{0pt}{$\mathbf{L}_{\text{best},i} \leftarrow \underset{\mathbf{X}_i \in \mathbf{N}_i}{\mathrm{argmin}} f(\mathbf{X}_i)$ for $i = 1, \hdots, S$ \textbf{//} current best design in local neighborhoods}
		\vspace{0.2cm}
		\STATE \hspace{0pt}{\textbf{//} swarm search loop}
		\STATE \hspace{0pt}{\textbf{while} stopping criteria not met \textbf{do} (this is iteration over $t$)} \label{search:loop:2}
		\STATE \hspace{8pt}{\textbf{if} $\mathbf{G}_{\text{best}}(t) = \mathbf{G}_{\text{best}}(t-1)$}
		\STATE \hspace{16pt}{\textbf{//} if the solution doesn't improve, reform the local communication networks}
		\STATE \hspace{16pt}{$\mathbf{N}_i \leftarrow \texttt{genNeighbors}(\mathbf{X}_i)$ }
		\STATE \hspace{8pt}{\textbf{endif}}
		\STATE \hspace{8pt}{\textbf{for each }{$i = 1, \ldots, S$ \textbf{do}} \textbf{//} Update velocities and positions} 
		\STATE \hspace{16pt}{$\mathbf{V}_i \leftarrow \omega \mathbf{V_i} + c_1 \mathbf{U} \odot\left(\mathbf{P}_{\text{best},i} - \mathbf{X}_i \right) + c_2\mathbf{U} \odot\left(\mathbf{L}_{\text{best},i}  - \mathbf{X}_i \right)$}
		\STATE \hspace{16pt}{$\{v_{ijk}\leftarrow\min\{v_{ijk}, v_{k}^\text{max}\}\}$ for $j = 1, \hdots, N$}
		\STATE \hspace{16pt}{$\mathbf{X}_i \leftarrow \mathbf{X}_i + \mathbf{V}_i$}
		\STATE \hspace{16pt}{$\mathbf{X}_i \leftarrow \texttt{confine} (\mathbf{X}_i)$ \textbf{//} keep candidates in searchspace}
		\STATE \hspace{16pt}{\textbf{if} $f(\mathbf{X}_i) < f(\mathbf{P}_{\text{best},i})$ \textbf{//} update knowledge about best known design to time $t$} 
		\STATE \hspace{24pt}{$\mathbf{P}_{\text{best},i} \leftarrow \mathbf{X}_i$}
		\STATE \hspace{24pt}{\textbf{if} $f(\mathbf{P}_{\text{best},i}) < f(\mathbf{G}_{\text{best}})$}
		\STATE \hspace{32pt}{$\mathbf{G}_{\text{best}} \leftarrow \mathbf{P}_{\text{best},i}$}
		\STATE \hspace{24pt}{\textbf{endif}}
		\STATE \hspace{16pt}{\textbf{endif}}
		\STATE \hspace{8pt}\textbf{endfor}
		\STATE \hspace{8pt}{\textbf{for each }{$i = 1, \ldots, S$ \textbf{do}} \textbf{//} with personal best's update, can update local best}
		\STATE \hspace{16pt}{\textbf{if} $f(\mathbf{P}_{\text{best},i}) < f(\mathbf{L}_{\text{best},i})$} 
		\STATE \hspace{24pt}{$\mathbf{L}_{\text{best},i} \leftarrow \mathbf{P}_{\text{best},i}$ \textbf{//} update best known design in local neighborhoods}
		\STATE \hspace{16pt}{\textbf{endif}}
		\STATE \hspace{8pt}\textbf{endfor}
		\STATE \hspace{0pt}{\textbf{endwhile}}
		\vspace{0.2cm}
    \STATE \textbf{Output:} Particle swarm solution---the best optimal design found $\mathbf{G}_{\text{best}}$
  \end{algorithmic}
\end{algorithm} 

\subsection{Computing Architecture}
\singlespace
We implemented Algorithm \ref{alg:bpso} in the \texttt{Julia} language \citep{julia}. Our initial prototyping was done in \texttt{R} \citep{R}, and while single runs of PSO were reasonably fast for low dimension problems, we identified the need for a faster implementation so that we could study several problems of moderate to high dimension in a reasonable time. \texttt{Julia} offers such speed on two fronts 1.) it is a high level syntax data science language, like \texttt{R}, \MATLAB, or \texttt{Python} and so offers the same rapid development speed as those languages, and 2.) unlike \texttt{R}, \MATLAB, or \texttt{Python}, which are interpreted languages, \texttt{Julia} is \emph{just in time compiled} meaning that the second run of the algorithm in \texttt{Julia} offers speeds on par with compiled languages such as C++.  Further, \texttt{Julia} has parallel computing inherently `built in' and so made it easy to run many independent runs of PSO on different computing cores of our CPU. Our machine is an Intel i7-6700K which has 4 cores and 8 compute threads running at 4.0GHz. Thus we could send multiple runs of PSO to 7 of the cores at the same time. 

\cite{gIopt} is the only reference that reported computing cost among our references, and they offer two measures: 1.) number of function evaluations over all runs for each design scenario and 2.) computing time wall-clock for the set of searches for each algorithm and design scenario and so a way to check the efficiency of $G$-PSO relative to the state-of-the-art. We will compare the efficiency of PSO to the other algorithms by reporting number of function evaluations during the runs of PSO. This measure is, of course, senstive to the number of particles and the stopping criterion. We ran all PSO searches with $S=150$ particles and our stopping criterion was a non-zero change in $f$ less than the square-root of machine-epsilon (about $10^{-8}$).

\subsection{Design Scenarios}
We ran $G$-PSO $n_{\text{run}} = 140$  times for all design scenarios (and searches for the optimal design that supports the second-order model) covered by \cite{jobo1}. These scenarios cover $K = 1, 2, 3$ design factors with experiment sizes $N = 3, 4, 5, 6, 7, 8, 9$, $N = 6, 7, 8, 9, 10, 11, 12$, and $N = 10, 11, 12, 13, 14, 15, 16$, respectively. Both \cite{jobo1} and \cite{gIopt} ran the GA for these scenarios, but, they did not find the same exact  $G$-optimal designs (in part related to the stopping criterion implemented by \cite{gIopt}). Regarding $G$-optimality, \cite{jobo1} found the  current best-known $G$-optimal designs (to this point). \cite{gIopt} found highly $G$-optimal designs relative to those of \cite{jobo1}, but used their GA generated designs to compare their results from the other two algorithms in their study. 

We also ran $G$-PSO $n_{\text{run}} = 210$ for the additional scenarios in \cite{rodman}. These scenarios cover $K = 4, 5$ with experiment sizes $N = 15, 20, 24$ and $N = 21, 26, 30$ respectively. Last, we ran $G$-PSO $n_{\text{run}} = 210$ for the additional scenarios in \cite{gIopt}, specifically $K = 4, N =17$ and $K = 5, N = 23$.

In total we've covered all published exact $G$-optimal designs for 29 design scenarios requiring 4620 independent runs of PSO. In the next section we compare $G$-PSO results to those of \cite{jobo1}, \cite{rodman}, and \cite{gIopt}.

\section{Results}
\label{sec:verify}

\subsection{The $K = 1, 2, 3$ design scenarios}
\begin{figure}
\begin{center}
  \includegraphics[width = \textwidth]{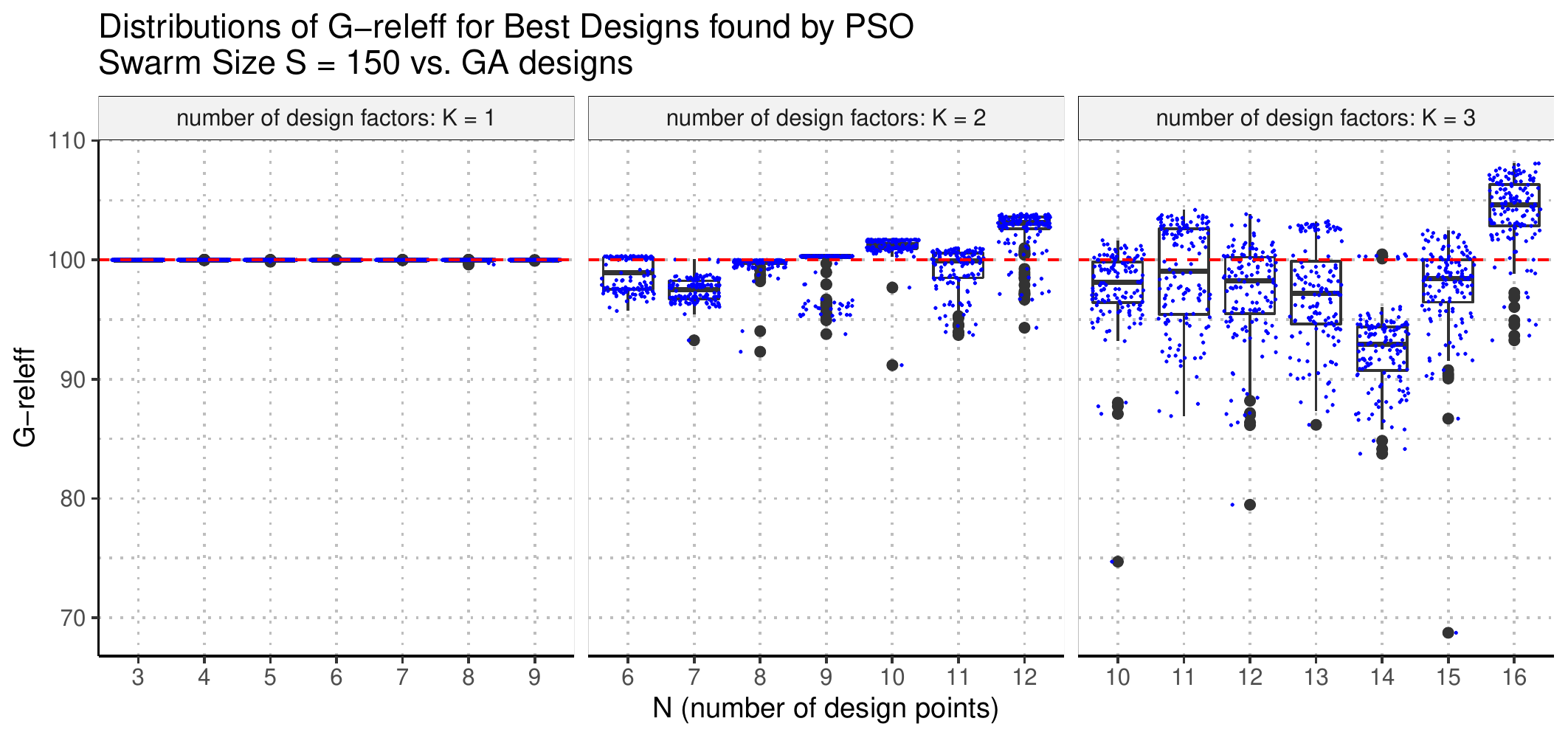}
\end{center}
\caption{Distributions of $G$-PSO design efficiencies relative to the published GA designs \label{fig:first}}
\end{figure}

The proposed $G$-optimal designs generated by the GA implemented by \cite{jobo1} were reproduced by $G$-CEXCH in \cite{rodman} and heretofore have remained the best known exact $G$-optimal designs for the second-order model and each of the $K=1,2,3$ design scenarios. We provide a comparison of the $n_{\text{run}}=140$ PSO search results (for each scenario) to the $G$-GA designs of \cite{jobo1} in Figure \ref{fig:first}. The data presented in this graphic are efficiencies of the $G$-PSO designs relative to the $G$-GA designs. Therefore, scores over 100 indicate that the PSO generated design is an improvement over the GA generated design. For the $K = 1$ scenarios, which are low-dimensional optimization from 3 to 9 dimensions, the graphic illustrates that all PSO runs are finding designs with equivalent $G$-score relative to the GA designs. In the second panel the results for the $K=2$ designs now shows a distribution in the $G$-releff score, but, for each scenario, the PSO produced designs with 90\% relative efficiency or higher for every single run. Further, PSO found a better $G$-optimal design for design sizes $N = 9, 10, 11, 12$. The last panel of the graphic indicates that, for $K=3$ factors and all experiment sizes $N$, PSO has identified an improved $G$-optimal design than those currently known, and the improvements are non-negligible, with increases ranging from 2-8\% efficiency. 

\citet{gIopt} provides a comparison of $G(I_\lambda)$-CEXCH, $G$-CEXCH and $G$-GA and their relative ability to generate designs with the highest/best $G$-optimality. We note that, in their data, \cite{gIopt} did not compare their results to the $G$-GA designs as published by \cite{jobo1}, rather, they re-ran the GA algorithm $n_{\text{run}} = 200$ times in order to compare relative costs of the algorithm. Thus, the $G$-GA optimal designs in \cite{gIopt} are equal to or less efficient than those of \cite{jobo1}. Nonetheless, we can use their data to compare all algorithms vs. the GA designs of \cite{jobo1} (which are the best known to date). Table \ref{tab:Geffs} reports the efficiencies of the optimal design relative to the GA design found by \cite{jobo1} for each scenario and algorithm. In all cases, it can be seen that PSO finds the best design vs. all other algorithms. 

Regarding computing cost, run time wall-clock for the entire set $140*21 = 2940$ of PSO searches over $K=1,2,3$ factors and the $N$ aforementioned was approximately 30 minutes (recall we are using \texttt{Julia} as well as parallel computing). \citet{gIopt} provides algorithm cost in the form of number of function evaluations over $n_{\text{run}} = 200$ runs of $G(I_\lambda)$-CEXCH, $G$-CEXCH and $G$-GA. Table \ref{tab:run} contains comparison of the \cite{gIopt} cost data to the cost of running $G$-PSO over our $n_{\text{run}} = 140$ runs. We provide a comparison of $G$-PSO cost on \cite{gIopt}'s $n_{\text{run}} = 200$ run scale by estimating the expected number of function evaluations and a 95\% confidence interval via Poisson statistics. Table \ref{tab:run} shows that $GA$ is the most expensive algorithm, and often via 2 orders of magnitude for the higher-dimensional problems. $G(I\lambda)$-CEXCH is slightly more costly than $G$-CEXCH for the lower dimensional problems, but is cheaper for the higher dimensional problems (often by an order of magnitude). Last, the 95\% confidence interval on the expected number of function evaluations in 200 PSO searches indicates that, in all cases, $G$-PSO has approximately the same cost as the new $G(I_\lambda)$-CEXCH (with some scenarios being slightly higher, but many scenarios being slightly lower cost than $G(I_\lambda)$-CEXCH). 

These results illustrate that $G$-PSO is approximately the same cost as the state-of-the-art algorithm $G(I_\lambda)$-CEXCH for generating $G$-optimal designs, while PSO generates highly optimal designs more efficiently, as it is demonstrated here to produce the current best-known exact $G$-optimal designs for these scenarios.

\begin{table}[]
\caption{Relative efficiencies of $G$-optimal designs for each algorithm relative to GA generated designs of \cite{jobo1}.} 
\label{tab:Geffs} 
\begin{center}
\begin{tabular}{ c c c c }
\cline{1-4}
\multicolumn{4}{c}{$K=1$ Factor}                                                                                   \\ \hline \hline
\multicolumn{1}{c|}{$N$} & \multicolumn{1}{c|}{$G$-CEXCH} & \multicolumn{1}{c|}{$G(I_{\lambda})$-CEXCH} & $G$-PSO  \\ \cline{1-4}
\multicolumn{1}{c|}{3}   & \multicolumn{1}{c|}{100.0}     & \multicolumn{1}{c|}{100.0}                  & 100.0    \\
\multicolumn{1}{c|}{4}   & \multicolumn{1}{c|}{98.7}      & \multicolumn{1}{c|}{96.2}                   & 100.0    \\
\multicolumn{1}{c|}{5}   & \multicolumn{1}{c|}{98.7}      & \multicolumn{1}{c|}{97.0}                   & 100.0    \\
\multicolumn{1}{c|}{6}   & \multicolumn{1}{c|}{100.0}     & \multicolumn{1}{c|}{100.0}                  & 100.0    \\
\multicolumn{1}{c|}{7}   & \multicolumn{1}{c|}{99.7}      & \multicolumn{1}{c|}{98.8}                   & 100.0    \\
\multicolumn{1}{c|}{8}   & \multicolumn{1}{c|}{99.4}      & \multicolumn{1}{c|}{94.7}                   & 100.0    \\
\multicolumn{1}{c|}{9}   & \multicolumn{1}{c|}{89.4}      & \multicolumn{1}{c|}{100.0}                  & 100.0    \\ \cline{1-4}
                         &                                &                                             &          \\ \cline{1-4}
\multicolumn{4}{c}{$K=2$ Factors}                                                                                  \\ \hline \hline
\multicolumn{1}{c|}{$N$} & \multicolumn{1}{c|}{$G$-CEXCH} & \multicolumn{1}{c|}{$G(I_{\lambda})$-CEXCH} & $G$-PSO  \\ \cline{1-4}
\multicolumn{1}{c|}{6}   & \multicolumn{1}{c|}{96.5}      & \multicolumn{1}{c|}{94.1}                   & 100.3    \\
\multicolumn{1}{c|}{7}   & \multicolumn{1}{c|}{97.9}      & \multicolumn{1}{c|}{95.5}                   & 100.1    \\
\multicolumn{1}{c|}{8}   & \multicolumn{1}{c|}{99.7}      & \multicolumn{1}{c|}{94.7}                   & 100.0    \\
\multicolumn{1}{c|}{9}   & \multicolumn{1}{c|}{97.0}      & \multicolumn{1}{c|}{95.8}                   & 100.3    \\
\multicolumn{1}{c|}{10}  & \multicolumn{1}{c|}{97.5}      & \multicolumn{1}{c|}{93.2}                   & 101.7    \\
\multicolumn{1}{c|}{11}  & \multicolumn{1}{c|}{94.0}      & \multicolumn{1}{c|}{97.0}                   & 101.0    \\
\multicolumn{1}{c|}{12}  & \multicolumn{1}{c|}{101.2}     & \multicolumn{1}{c|}{95.1}                   & 103.9    \\ \cline{1-4}
                         &                                &                                             &          \\ \cline{1-4}
\multicolumn{4}{c}{$K=3$ Factors}                                                                                  \\ \hline \hline
\multicolumn{1}{c|}{$N$} & \multicolumn{1}{c|}{$G$-CEXCH} & \multicolumn{1}{c|}{$G(I_{\lambda})$-CEXCH} & $G$-PSO  \\ \cline{1-4}
\multicolumn{1}{c|}{10}  & \multicolumn{1}{c|}{93.1}      & \multicolumn{1}{c|}{95.4}                   & 101.6    \\
\multicolumn{1}{c|}{11}  & \multicolumn{1}{c|}{92.9}      & \multicolumn{1}{c|}{96.9}                   & 104.2    \\
\multicolumn{1}{c|}{12}  & \multicolumn{1}{c|}{90.7}      & \multicolumn{1}{c|}{90.3}                   & 103.8    \\
\multicolumn{1}{c|}{13}  & \multicolumn{1}{c|}{92.9}      & \multicolumn{1}{c|}{99.9}                   & 103.2    \\
\multicolumn{1}{c|}{14}  & \multicolumn{1}{c|}{87.6}      & \multicolumn{1}{c|}{100.0}                  & 100.5    \\
\multicolumn{1}{c|}{15}  & \multicolumn{1}{c|}{98.5}      & \multicolumn{1}{c|}{100.1}                  & 102.5    \\
\multicolumn{1}{c|}{16}  & \multicolumn{1}{c|}{100.1}     & \multicolumn{1}{c|}{100.2}                  & 108.1    \\ \cline{1-4}
\end{tabular}
\end{center}
\end{table}

\begin{landscape}
\begin{table}[]
\caption{Algorithm cost comparison. Table value is log10(\# $f$ evaluations). The first PSO column reports the observed number of function evaluations in 140 runs of PSO. The gray columns report an estimate of the expected number of function evaluations in 200 PSO runs with a 95\% confidence interval.}
\label{tab:run}
\begin{center}
\renewcommand{\arraystretch}{1.15}
\begin{tabular}{ccc|ccccc
}
\hline\hline
\multicolumn{2}{c}{Design Scenario} & \multicolumn{3}{|c|}{$G$-PSO}                                                                      & \multicolumn{1}{c}{$G(I_\lambda)$-CEXCH} & \multicolumn{1}{c}{$G$-CEXCH} & $G$-GA \\ \hline
$K$                        & \multicolumn{1}{c}{$N$}                 & \multicolumn{1}{|c|}{\cellcolor[HTML]{FFFFFF}$n_{\text{run}} = 140$} &  \multicolumn{2}{c}{$n_{\text{run}} = 200$}    & \multicolumn{3}{c}{$n_{\text{run}} = 200$}                                                                                                                                                \\ \hline\hline
\textbf{}                  & \textbf{}                                & \textbf{}                                                      & \textbf{estimate}                & \textbf{95\% CI}               & \textbf{}                                                                  &                                                                 &                                         \\
1                          & 3                                        & 6.000                                                          & 6.155                            & (6.145, 6.165)                 & 6.0                                                                        & 5.5                                                             & 6.9                                     \\
1                          & 4                                        & 6.535                                                          & 6.690                             & (6.684, 6.695)                 & 6.4                                                                        & 5.7                                                             & 7.0                                     \\
1                          & 5                                        & 6.681                                                          & 6.835                            & (6.831, 6.840)                 & 6.6                                                                        & 5.8                                                             & 7.1                                     \\
1                          & 6                                        & 6.226                                                          & 6.381                            & (6.373, 6.388)                 & 6.4                                                                        & 5.9                                                             & 7.2                                     \\
1                          & 7                                        & 6.685                                                          & 6.840                             & (6.835, 6.845)                 & 6.8                                                                        & 6.0                                                             & 7.2                                     \\
1                          & 8                                        & 6.761                                                          & 6.916                            & (6.912, 6.921)                 & 6.8                                                                        & 6.0                                                             & 7.3                                     \\
1                          & 9                                        & 6.405                                                          & 6.560                             & (6.553, 6.566)                 & 6.9                                                                        & 6.1                                                             & 7.4                                     \\
                           &                                          &                                                                &                                  &                                &                                                                            &                                                                 &                                         \\
2                          & 6                                        & 7.088                                                          & 7.243                            & (7.240, 7.246)                 & 7.2                                                                        & 7.3                                                             & 8.4                                     \\
2                          & 7                                        & 7.086                                                          & 7.241                            & (7.238, 7.244)                 & 7.4                                                                        & 7.3                                                             & 8.5                                     \\
2                          & 8                                        & 7.042                                                          & 7.197                            & (7.194, 7.200)                 & 7.2                                                                        & 7.5                                                             & 8.6                                     \\
2                          & 9                                        & 7.119                                                          & 7.274                            & (7.271, 7.277)                 & 7.0                                                                        & 7.5                                                             & 8.6                                     \\
2                          & 10                                       & 7.163                                                          & 7.318                            & (7.315, 7.321)                 & 7.3                                                                        & 7.6                                                             & 8.7                                     \\
2                          & 11                                       & 7.221                                                          & 7.376                            & (7.373, 7.378)                 & 7.4                                                                        & 7.6                                                             & 8.7                                     \\
2                          & 12                                       & 7.196                                                          & 7.351                            & (7.348, 7.354)                 & 7.7                                                                        & 7.7                                                             & 8.7                                     \\
                           &                                          &                                                                &                                  &                                &                                                                            &                                                                 &                                         \\
3                          & 10                                       & 7.437                                                          & 7.592                            & (7.590, 7.594)                 & 8.0                                                                        & 8.7                                                             & 9.6                                     \\
3                          & 11                                       & 7.511                                                          & 7.666                            & (7.664, 7.668)                 & 7.8                                                                        & 8.8                                                             & 9.7                                     \\
3                          & 12                                       & 7.544                                                          & 7.699                            & (7.697, 7.701)                 & 7.9                                                                        & 8.8                                                             & 9.7                                     \\
3                          & 13                                       & 7.538                                                          & 7.692                            & (7.691, 7.694)                 & 7.5                                                                        & 8.9                                                             & 9.7                                     \\
3                          & 14                                       & 7.543                                                          & 7.698                            & (7.696, 7.700)                 & 7.6                                                                        & 9.0                                                             & 9.8                                     \\
3                          & 15                                       & 7.515                                                          & 7.670                             & (7.668, 7.671)                 & 7.6                                                                        & 9.2                                                             & 9.8                                     \\
3                          & 16                                       & 7.556                                                          & 7.711                            & (7.709,  7.713)                & 7.6                                                                        & 9.9                                                             & 9.8                                     \\ \hline
\end{tabular}
\end{center}

\end{table}
\end{landscape}

\subsection{The $K = 4, 5$ design scenarios}
Due to the computational cost, this number of experimental factors is the highest that the design community has gone to date. Searching for $G$-optimal designs for more factors will take a considerable time/computing investment. Note, however, that $K$ does not define the dimension of the optimization search: the dimension is $N*K$ so the largest problem we study here is $K = 5$, $N = 30$ which is an optimization search in a 150 dimensional parameter space. 

\begin{figure}
\begin{center}
  \includegraphics[width = \textwidth]{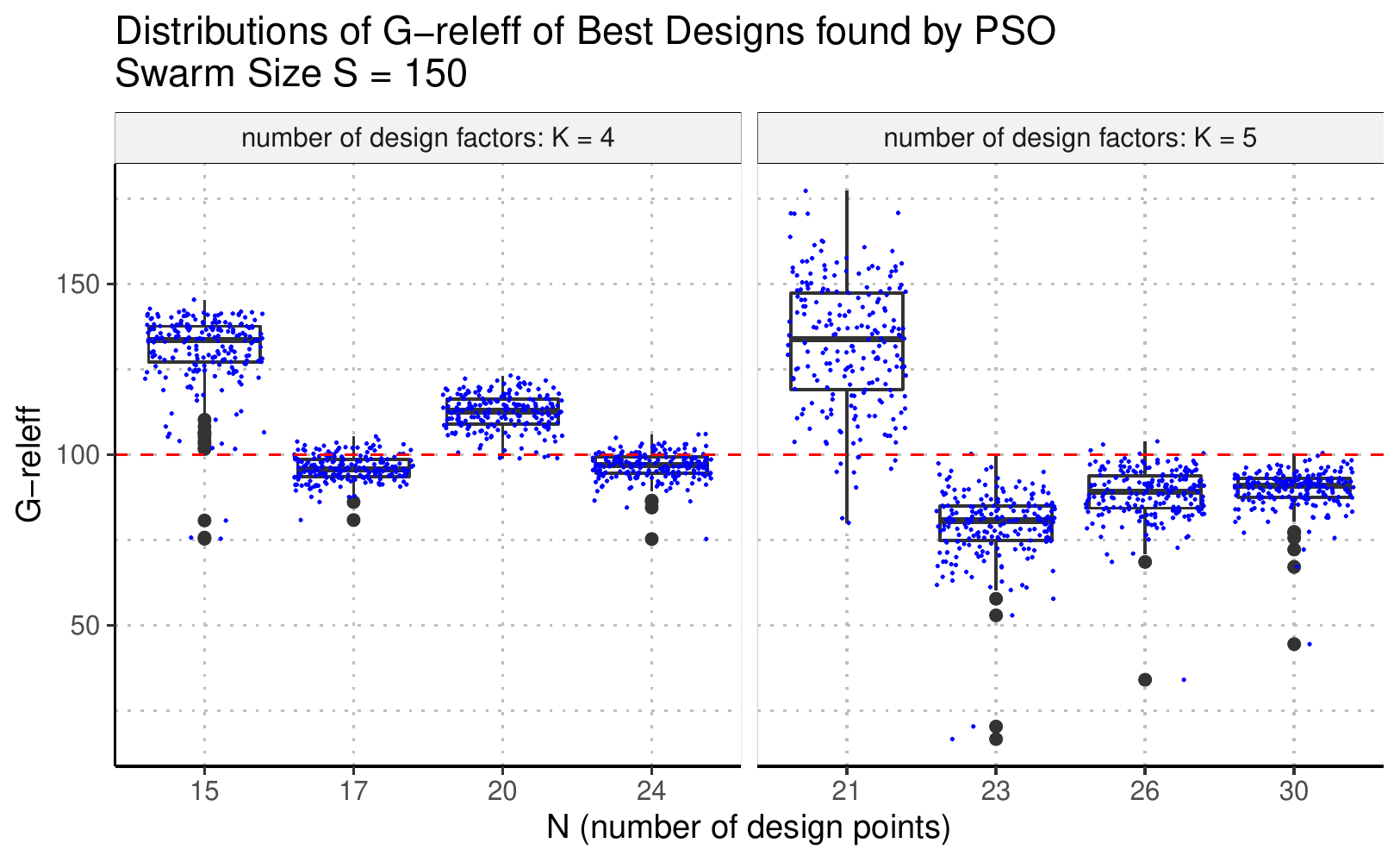}
\end{center}
\caption{Distributions of $G$-PSO design efficiencies relative to the published CEXCH designs for $K=4, 5$ factors.\label{fig:sec}}
\end{figure}

Table \ref{tab:new45} contains the $G$-efficiencies of the best $G(I_\lambda)$-CEXCH, $G$-CEXCH, and $G$-PSO generated design, as well as the relative efficiency of the $G$-PSO design to the indicated CEXCH algorithm. In all cases PSO is found to generate better $G$-optimal designs, and in some cases with a significant improvement. We present distributions of the $G$-releff scores of the PSO designs to the corresponding CEXCH generated designs in Figure \ref{fig:sec}. In all cases it can be seen that PSO found an equivalent or better $G$-optimal design (evidenced by relative efficiencies over 100). We note that each of the CEXCH algorithms was run $n_{\text{run}}=200$ times in the work of \cite{gIopt} and the PSO searches were run $n_{\text{run}}=210$ times (a number evenly distributed on 7 computer cores). The graphic further illustrates PSO's ability to seek highly optimal designs each run, evidenced by the distributions of $G$-releff being tightly packed at or over 100\% relative efficiency. For many scenarios, there is apparently a high probability that PSO would generate a design with 95\% efficiency or better in a single run.

Regarding the significant improvements in designs, 1.) for the $K=4, N = 15$ case PSO provided a design with 145\% relative efficiency to the best-known design, 2.) for the $K = 4, N = 20$ scenario, PSO produced a design with 123\% improved efficiency, and 3.) for the $K=5, N = 21$ scenario, PSO found a design with 177\% relative efficiency. We contacted the authors of \cite{rodman} to investigate these large discrepancies. For the $K=4, N = 15$ and $K = 5, N = 21$ scenarios, it was confirmed that the designs published by \cite{rodman} were mis-scored on the $G$-scale (personal email correspondence with Dr. Bradley Jones, June 20, 2020). Given that \cite{rodman} employs a separate optimization search to score each candidate design on the $G$-scale, these results illustrate the consequences of failing to find the maximum prediction variance for a candidate design, and, to our opinion, support the approach of using the $5^K$ grid $\mathbf{G}_\mathcal{X}$ to score candidate designs. 

The information for a proper time comparison for generating designs with $K=4, 5$ factors via PSO vs. the other approaches does not exist due to algorithms being run on different machines and computing languages. Nonetheless, for information we report what data do exist for computing times on the $K=4, N = 17$ and $K=5, N = 23$ design scenarios. \citet{gIopt} report that they were able to run $G(I_\lambda)$-CEXCH $n_{\text{run}} = 200$ times on the $K=4, N = 17$ in 20.13 hours on their computing platform (approx 6.0m for each run). They were not able to run $G$-CEXCH as they estimated that 200 runs would have taken 25 days on their machine. Our approach (PSO, Julia, $n_{\text{run}} = 210$ parallel PSO runs [30 runs per 7 cores]) took about 3 hours which translates into approximately 6m per each individual run. \citet{gIopt} report that they were able to run $G(I_\lambda)$-CEXCH $n_{\text{run}} = 200$ times on the $K=5, N = 23$ in 25.07 hours on their computing platform (approx 7.5m for each run). They were not able to run $G$-CEXCH as they estimated that 200 runs would have taken 166 days on their machine. Our approach (PSO, Julia, $n_{\text{run}} = 210$ parallel PSO runs [30 runs per 7 cores]) took about 20h which translates into approximately 40m per each individual run. These results imply that PSO has more difficulty scaling to higher dimension. The reason why PSO took approximately 6 times longer for the $K=5$ scenario than for the $K=4$ scenario is that the respective $5^K$ grids used to score each candidate design during the search have $5^5 = 3125$ and $5^4 = 625$ points respectively, and \emph{each particle} (\ie each candidate design) must have the $G$-score evaluated \emph {at each of these grid points, at each iteration of the algorithm}. Nonetheless, we believe this time increase to be of little hindrance to scaling PSO to efficient searches for higher dimensional designs due to the use of Julia and parallel computing, and the additional cost is easily mitigated via CPUs with more computer cores.

\begin{table}[]
\caption{$G$-efficiencies of published $G$-optimal designs for $K=4,5$ factors. The last column is the relative efficiency of the best $G$-PSO design compared to previously published designs. Author \cite{rodman} used the $G$-CEXCH while \cite{gIopt} used the $G(I_\lambda)$-CEXCH.} 
\label{tab:new45} 
\begin{center}
\begin{tabular}{ccccc}
\hline
\multicolumn{5}{c}{$K=4$ Factors}                                                                                                         \\ \hline \hline
\multicolumn{1}{c|}{$N$} & \multicolumn{1}{c|}{source}           & \multicolumn{1}{c|}{CEXCH} & \multicolumn{1}{c|}{$G$-PSO} & $G$-releff \\ \hline
\multicolumn{1}{c|}{15}  & \multicolumn{1}{c|}{Rodriguez (2010)} & \multicolumn{1}{c|}{48.89} & \multicolumn{1}{c|}{71.09}   & 145.41     \\
\multicolumn{1}{c|}{17}  & \multicolumn{1}{c|}{Hernandez (2018)} & \multicolumn{1}{c|}{70.14} & \multicolumn{1}{c|}{73.90}   & 105.36     \\
\multicolumn{1}{c|}{20}  & \multicolumn{1}{c|}{Rodriguez (2010)} & \multicolumn{1}{c|}{65.11} & \multicolumn{1}{c|}{80.20}   & 123.18     \\
\multicolumn{1}{c|}{24}  & \multicolumn{1}{c|}{Rodriguez (2010)} & \multicolumn{1}{c|}{81.05} & \multicolumn{1}{c|}{85.95}   & 106.05     \\ \hline
                         &                                       &                            &                              &            \\ \hline
\multicolumn{5}{c}{$K=5$ Factors}                                                                                                         \\ \hline \hline
\multicolumn{1}{c|}{$N$} & \multicolumn{1}{c|}{source}           & \multicolumn{1}{c|}{CEXCH} & \multicolumn{1}{c|}{$G$-PSO} & $G$-releff \\ \hline
\multicolumn{1}{c|}{21}  & \multicolumn{1}{c|}{Rodriguez (2010)} & \multicolumn{1}{c|}{38.74} & \multicolumn{1}{c|}{68.67}   & 177.26     \\
\multicolumn{1}{c|}{23}  & \multicolumn{1}{c|}{Hernandez (2018)} & \multicolumn{1}{c|}{73.02} & \multicolumn{1}{c|}{73.19}   & 100.24     \\
\multicolumn{1}{c|}{26}  & \multicolumn{1}{c|}{Rodriguez (2010)} & \multicolumn{1}{c|}{72.47} & \multicolumn{1}{c|}{75.31}   & 103.92     \\
\multicolumn{1}{c|}{30}  & \multicolumn{1}{c|}{Rodriguez (2010)} & \multicolumn{1}{c|}{75.80} & \multicolumn{1}{c|}{76.16}   & 100.47     \\ \hline
\end{tabular}
\end{center}
\end{table}

\section{Conclusion}
\label{sec:conc}
In this paper we summarized the last two decades of research into algorithm development for generating exact $G$-optimal designs. To date, PSO has not been applied to this problem and so we proposed an extension of PSO to generate exact $G$-optimal designs. We then ran PSO on all published design scenarios. For the 21 $N$ scenarios on $K=1, 2, 3$ design factors studied in \cite{jobo1, rodman, gIopt}, PSO found better $G$-optimal designs for 12 of the higher dimension scenarios, some of these improvements offering 5\% or better efficiency than currently known designs. On our computing setup, the total run time for all 2940 PSO searches on the 21 $K=1,2,3$ design scenarios was less than 30 minutes with the the $n_{\text{run}}=140$ PSO searches for the $K=1$ scenarios taking less than a minute, and the $n_{\text{run}}=140$ PSO searches for the $K= 3$, $N = 16$ (\ie a 48 dimensional optimization) took approximately 4 minutes, or about 12 seconds for each individual PSO search (using $S=150$ particles). The use of \texttt{Julia} and parallel computing definitely helps to run many PSO searches very quickly. We believe this speed may enable realistic searches for good $G$-optimal designs for practitioners.

There are not many published proposed exact $G$-optimal designs for $K = 4, 5$ design factor cases due to the expense of searching for optimal designs for these scenarios. To date, there are only 4 design scenarios for each $K$ covered by \cite{rodman} and \cite{gIopt}. In all cases, $G$-PSO found as good or better $G$-optimal designs than those currently known. 

PSO is distinct from the various coordinate exchange algorithms studied in the following way. CEXCH starts with a randomly drawn design and then optimizes this design locally in $\mathcal{X}^N$. In CEXCH, there is no `intelligence' which seeks to search $\mathcal{X}^N$ more globally for the best possible optimum. PSO, in contrast, uses $S$ randomly drawn design matrices which are searching $\mathcal{X}^N$ for the best possible fitness on the objective. Further, these candidate designs remember and communicate their best positions, have a tendency to want to revisit these locations, and this increases the likelihood of PSO to find the global optimum.  In this sense, and due to the demonstrated computational cost, we propose that PSO should now be viewed as state-of-the art for generating optimal designs.

The genetic algorithm has enjoyed large success in academic research and is demonstrated, to this point, to be the superior algorithm for generating optimal designs with the best optimality scores. The GA, however, is computationally expensive, and so it is not a great tool for the practitioner of experiments to generate candidate optimal designs for their problem. This work has demonstrated that PSO, in repeated runs, performs as well or better than GA in generating optimal designs at a fraction of the cost. Therefore, we propose that PSO should be applied further in academic research in generating optimal designs instead of GA.

Our future research trajectory in this regard is as follows. We will extend our PSO code to generate designs for linearly constrained design spaces. We will also apply PSO to generating other optimal designs \eg $D$-, $I$-, and $A$- designs are popular to study in research---as $G$-optimal design generation is a much harder problem than these other criteria, we are expecting PSO to perform well. We are also planning to apply PSO to generate optimal designs with a specified pure-replication structure. Last, we plan to extend the PSO to generating optimal designs for mixture experiments. The results presented in this paper suggest PSO will perform well in these future applications.

In conclusion, PSO is hereby demonstrated to be superior to existing algorithms for generating exact $G$-optimal designs. It costs roughly the same as the state-of-the-art algorithm, but is more efficient at finding the global optimal design in all studied cases.

\bigskip
\begin{center}
{\large\bf SUPPLEMENTARY MATERIAL}
\end{center}

\begin{description}
\item[PSO $G$-optimal designs:] File containing all PSO generated $G$-optimal designs for all scenarios discussed in this paper (can be used to verify optimality scores). (.csv file)
\item[R-code for scoring the newly found proposed exact $G$-optimal designs:] this script may be applied to the provided .csv file in order to verify the $G$-scores of the PSO generated designs published in this paper, and to reproduce several of the discussed results. 
\end{description}

\bibliographystyle{spbasic}      
\bibliography{mybib}   

\begin{thebibliography}{34}
\providecommand{\natexlab}[1]{#1}
\providecommand{\url}[1]{{#1}}
\providecommand{\urlprefix}{URL }
\expandafter\ifx\csname urlstyle\endcsname\relax
  \providecommand{\doi}[1]{DOI~\discretionary{}{}{}#1}\else
  \providecommand{\doi}{DOI~\discretionary{}{}{}\begingroup
  \urlstyle{rm}\Url}\fi
\providecommand{\eprint}[2][]{\url{#2}}

\bibitem[{Atkinson and Bailey(2001)}]{atkRev}
Atkinson A, Bailey R (2001) One hundred years of the design of experiments on
  and off the pages of biometrika. Biometrika 88(1):53--97

\bibitem[{Bezanson et~al.(2017)Bezanson, Edelman, Karpinski, and Shah}]{julia}
Bezanson J, Edelman A, Karpinski S, Shah VB (2017) Julia: A fresh approach to
  numerical computing. SIAM review 59(1):65--98,
  \urlprefix\url{https://doi.org/10.1137/141000671}

\bibitem[{Borkowski(2003)}]{jobo1}
Borkowski J (2003) Using a genetic algorithm to generate small exact response
  surface designs. Journal of Probability and Statistical Science 1(1)

\bibitem[{Bratton and Kennedy(2007)}]{spso1_2007}
Bratton D, Kennedy J (2007) Defining a standard for particle swarm
  optimization. 2007 IEEE Swarm Intelligence Symposium pp 120--127

\bibitem[{{Bratton} and {Kennedy}(2007)}]{bratton}
{Bratton} D, {Kennedy} J (2007) Defining a standard for particle swarm
  optimization. In: 2007 IEEE Swarm Intelligence Symposium, pp 120--127,
  \doi{10.1109/SIS.2007.368035}

\bibitem[{Chen et~al.(2011)Chen, Chang, Wang, and Wong}]{chen0}
Chen R, Chang SP, Wang W, Wong W (2011) Optimal experimental designs via
  particle swarm optimization methods. Tech. rep., Department of Mathematics,
  National Taiwan University

\bibitem[{Chen et~al.(2013)Chen, Hsieh, Hung, and Wang}]{chen2}
Chen RB, Hsieh DN, Hung Y, Wang W (2013) Optimizing latin hypercube designs by
  particle swarm. Statistics and Computing 23(5):663--676,
  \doi{10.1007/s11222-012-9363-3},
  \urlprefix\url{https://doi.org/10.1007/s11222-012-9363-3}

\bibitem[{{Chen} et~al.(2014){Chen}, {Hsu}, {Hung}, and {Wang}}]{chen1}
{Chen} RB, {Hsu} YW, {Hung} Y, {Wang} W (2014) Discrete particle swarm
  optimization for constructing uniform design on irregular regions.
  Computational Statistics and Data Analysis 72:282--297,
  \doi{10.1016/j.csda.2013.10.015}

\bibitem[{Chen et~al.(2015)Chen, Chang, Wang, Tung, and Wong}]{Chen3}
Chen RB, Chang SP, Wang W, Tung HC, Wong WK (2015) Minimax optimal designs via
  particle swarm optimization methods. Statistics and Computing 25(5):975--988,
  \doi{10.1007/s11222-014-9466-0},
  \urlprefix\url{https://doi.org/10.1007/s11222-014-9466-0}

\bibitem[{Chen et~al.(2019)Chen, Li, Hung, and Wang}]{chen4}
Chen RB, Li CH, Hung Y, Wang W (2019) Optimal noncollapsing space-filling
  designs for irregular experimental regions. Journal of Computational and
  Graphical Statistics 28(1):74--91, \doi{10.1080/10618600.2018.1482760},
  \urlprefix\url{https://doi.org/10.1080/10618600.2018.1482760},
  \eprint{https://doi.org/10.1080/10618600.2018.1482760}

\bibitem[{{Clerc}(1999)}]{clerc1}
{Clerc} M (1999) The swarm and the queen: towards a deterministic and adaptive
  particle swarm optimization. In: Proceedings of the 1999 Congress on
  Evolutionary Computation-CEC99 (Cat. No. 99TH8406), vol~3, pp 1951--1957 Vol.
  3, \doi{10.1109/CEC.1999.785513}

\bibitem[{Clerc(2012)}]{hal_spso}
Clerc M (2012) Standard particle swarm optimisation. Tech. rep., HAL
  achives-ouvertes

\bibitem[{{Clerc} and {Kennedy}(2002)}]{clerc2}
{Clerc} M, {Kennedy} J (2002) The particle swarm - explosion, stability, and
  convergence in a multidimensional complex space. IEEE Transactions on
  Evolutionary Computation 6(1):58--73, \doi{10.1109/4235.985692}

\bibitem[{{Eberhart} and {Kennedy}(1995)}]{opso2_1995}
{Eberhart} R, {Kennedy} J (1995) A new optimizer using particle swarm theory.
  In: MHS'95. Proceedings of the Sixth International Symposium on Micro Machine
  and Human Science, pp 39--43, \doi{10.1109/MHS.1995.494215}

\bibitem[{Eberhart and Shi(1998)}]{eberhart_1998}
Eberhart RC, Shi Y (1998) Comparison between genetic algorithms and particle
  swarm optimization. In: Porto VW, Saravanan N, Waagen D, Eiben AE (eds)
  Evolutionary Programming VII, Springer Berlin Heidelberg, Berlin, Heidelberg,
  pp 611--616

\bibitem[{{Eberhart} and {Shi}(2000)}]{ebershi_3}
{Eberhart} RC, {Shi} Y (2000) Comparing inertia weights and constriction
  factors in particle swarm optimization. In: Proceedings of the 2000 Congress
  on Evolutionary Computation. CEC00 (Cat. No.00TH8512), vol~1, pp 84--88
  vol.1, \doi{10.1109/CEC.2000.870279}

\bibitem[{{Engelbrecht}(2013)}]{engel}
{Engelbrecht} AP (2013) Particle swarm optimization: Global best or local best?
  In: 2013 BRICS Congress on Computational Intelligence and 11th Brazilian
  Congress on Computational Intelligence, pp 124--135,
  \doi{10.1109/BRICS-CCI-CBIC.2013.31}

\bibitem[{Fisher(1935)}]{fisher:1935}
Fisher R (1935) {The design of experiments. 1935}. Oliver and Boyd, Edinburgh

\bibitem[{Hernandez and Nachtsheim(2018)}]{gIopt}
Hernandez LN, Nachtsheim CJ (2018) Fast computation of exact {G}-optimal
  designs via {I}$_\lambda$-optimality. Technometrics 60(3):297--305,
  \doi{10.1080/00401706.2017.1371080},
  \urlprefix\url{https://doi.org/10.1080/00401706.2017.1371080},
  \eprint{https://doi.org/10.1080/00401706.2017.1371080}

\bibitem[{{Kennedy}(1997)}]{ken_1997}
{Kennedy} J (1997) The particle swarm: social adaptation of knowledge. In:
  Proceedings of 1997 IEEE International Conference on Evolutionary Computation
  (ICEC '97), pp 303--308, \doi{10.1109/ICEC.1997.592326}

\bibitem[{Kennedy(2006)}]{zombook}
Kennedy J (2006) Swarm Intelligence, Springer, pp 187--220

\bibitem[{{Kennedy} and {Eberhart}(1995)}]{opso_1995}
{Kennedy} J, {Eberhart} R (1995) Particle swarm optimization. In: Proceedings
  of ICNN'95 - International Conference on Neural Networks, vol~4, pp
  1942--1948 vol.4, \doi{10.1109/ICNN.1995.488968}

\bibitem[{Kiefer(1959)}]{kiefer}
Kiefer J (1959) Optimum experimental designs. Journal of the Royal Statistical
  Society Series B (Methodological) 21(2):272--319,
  \urlprefix\url{http://www.jstor.org/stable/2983802}

\bibitem[{Lukemire et~al.(2019)Lukemire, Mandal, and Wong}]{lukemire}
Lukemire J, Mandal A, Wong WK (2019) d-qpso: A quantum-behaved particle swarm
  technique for finding d-optimal designs with discrete and continuous factors
  and a binary response. Technometrics 61(1):77--87,
  \doi{10.1080/00401706.2018.1439405},
  \urlprefix\url{https://doi.org/10.1080/00401706.2018.1439405},
  \eprint{https://doi.org/10.1080/00401706.2018.1439405}

\bibitem[{Mak and Joseph(2018)}]{joseph}
Mak S, Joseph VR (2018) Minimax and minimax projection designs using
  clustering. Journal of Computational and Graphical Statistics 27(1):166--178,
  \doi{10.1080/10618600.2017.1302881},
  \urlprefix\url{https://doi.org/10.1080/10618600.2017.1302881},
  \eprint{https://doi.org/10.1080/10618600.2017.1302881}

\bibitem[{Meyer and Nachtsheim(1995)}]{mncoord}
Meyer RK, Nachtsheim CJ (1995) The coordinate-exchange algorithm for
  constructing exact optimal experimental designs. Technometrics 37(1):60--69,
  \doi{10.1080/00401706.1995.10485889},
  \urlprefix\url{https://www.tandfonline.com/doi/abs/10.1080/00401706.1995.10485889},
  \eprint{https://www.tandfonline.com/doi/pdf/10.1080/00401706.1995.10485889}

\bibitem[{{Myers} et~al.(2016){Myers}, {Montgomery}, and {Anderson-Cook}}]{rsm}
{Myers} R, {Montgomery} D, {Anderson-Cook} C (2016) Response surface
  methodology: process and product optimization using Designed Experiments. 4th
  Edition. John Wiley and Sons, Ltd.

\bibitem[{{R Core Team}(2019)}]{R}
{R Core Team} (2019) R: A Language and Environment for Statistical Computing. R
  Foundation for Statistical Computing, Vienna, Austria,
  \urlprefix\url{https://www.R-project.org/}

\bibitem[{Rodríguez et~al.(2010)Rodríguez, Jones, Borror, and
  Montgomery}]{rodman}
Rodríguez M, Jones B, Borror CM, Montgomery DC (2010) Generating and assessing
  exact g-optimal designs. Journal of Quality Technology 42(1):3--20,
  \doi{10.1080/00224065.2010.11917803},
  \urlprefix\url{https://doi.org/10.1080/00224065.2010.11917803},
  \eprint{https://doi.org/10.1080/00224065.2010.11917803}

\bibitem[{Shi and Eberhart(1998)}]{shi2}
Shi Y, Eberhart RC (1998) Parameter selection in particle swarm optimization.
  In: Porto VW, Saravanan N, Waagen D, Eiben AE (eds) Evolutionary Programming
  VII, Springer Berlin Heidelberg, Berlin, Heidelberg, pp 591--600

\bibitem[{Smith(1918)}]{firstGopt}
Smith K (1918) On the standard deviations of adjusted and interpolated values
  of an observed polynomial function and its constants and the guidance they
  give towards the proper choice of the distributions of observations.
  Biometrika XII(1 and 2)

\bibitem[{Walsh(2021)}]{walsh}
Walsh SJ (2021) Development and applications of particle swarm optimization for
  constructing optimal experimental designs. {PhD} dissertation, Montana State
  University,
  \urlprefix\url{https://drive.google.com/file/d/1KmFNQU75TJfQA6cPRksagHWlvbcuyPPB/view?usp=sharing}

\bibitem[{Wong et~al.(2015)Wong, Chen, Huang, and Wang}]{wong}
Wong WK, Chen RB, Huang CC, Wang W (2015) A modified particle swarm
  optimization technique for finding optimal designs for mixture models. PLOS
  ONE 10(6):1--23, \doi{10.1371/journal.pone.0124720},
  \urlprefix\url{https://doi.org/10.1371/journal.pone.0124720}

\bibitem[{Zambrano-Bigiarini et~al.(2013)Zambrano-Bigiarini, Clerc, and
  Rojas-Mujica}]{spso2_2011}
Zambrano-Bigiarini M, Clerc M, Rojas-Mujica R (2013) Standard particle swarm
  optimisation 2011 at cec-2013: A baseline for future pso improvements. 2013
  IEEE Congress on Evolutionary Computation pp 2337--2344

\end{thebibliography}

%
%

\end{document}